\documentstyle[eqsecnum,aps]{revtex}

\begin{document}
\draft \preprint{HEP/123-qed}
\title{Path Integral Computation of Phonon Anharmonicity }
\author{Marco Zoli}
\address{Istituto Nazionale Fisica della Materia - Universit\'a di Camerino, \\
62032 Italy. e-mail: zoli.marco@libero.it }

\date{\today}
\maketitle
\begin{abstract}
The partition function of an oscillator disturbed by a set of
electron particle paths has been computed by a path integral
method which permits to evaluate at any temperature the relevant
cumulant terms in the series expansion. The time dependent source
current peculiar of the semiclassical Su-Schrieffer-Heeger model
induces large electron-phonon anharmonicities on the phonon
subsystem. As a main signature of anharmonicity the phonon heat
capacity shows a peak whose temperature location strongly varies
with the strength of the {\it e-ph} coupling. High energy
oscillators are less sensitive to anharmonic perturbations.
\end{abstract} 

\narrowtext
\section*{I. Introduction}

There is at present a growing interest in electron-phonon non
linearities also triggered by the signatures of a large
anharmonicity observed in metal diborides \cite{cava}. Phonon
anharmonicities have a long story in solid state physics closely
related to the inelastic neutron scattering theory
\cite{weinstock,marad}. As the anharmonic effects are generally
small in crystals up to room temperature, second order
perturbation theory suffices to determine the lifetime due to
three phonons decay processes while the renormalization of the
frequencies, together with the three phonons terms, also requires
computation of the four phonons vertex in the first order diagram
plus a (dominant) contribution due to the thermal expansion of the
crystal \cite{io1}. It is known \cite{barron,allen,jarrell} that
most properties of real materials can be well described by
replacing the anharmonic phonons with the temperature dependent
renormalized harmonic phonons, that is assuming  quasi-harmonic
vibrational models \cite{io2}. Instead, the damping of some
anomalous bulk and surface phonons requires explicit computation
of the anharmonic interactions \cite{jay,io3}. However, neither
first principles calculations of anharmonicities based on density
functional perturbation theory \cite{bernardi} nor empirical force
constant approaches do separate the bare phonon-phonon
interactions from the non linearities due to the electron-phonon
coupling. The latter contribution is generally incorporated in the
former by fitting the third and fourth derivatives of the
interatomic potential to experimental thermoelastic properties. In
systems such as polymers, whose thermal and conducting behavior is
shaped by the strength of the e-ph coupling, one would desire to
estimate the amount of {\it e-ph} anharmonicity  which may become
relevant in the low temperature region where, instead, the
phonon-phonon interactions tend to vanish.

Theoretical investigations on polymers usually depart from the
Su-Schrieffer-Heeger (SSH) Hamiltonian \cite{ssh,raedt,ono} in
which the {\it e-ph} coupling is the derivative of the electron
hopping integral with respect to the intersite atomic
displacement. As the electron propagator couples to the oscillator
displacement the SSH model can be attacked by the path integral
method \cite{feynman,path,kleinert} which allows one to derive the
time dependent probability amplitude for a particle in a bath of
oscillators \cite{io4}. On the other hand, considering the
electron particle path as the disturbing source for the phonon
subsystem one may evaluate the amount of {\it e-ph} anharmonicity
by expanding perturbatively the phonon partition function.
Analysis of the cumulant terms versus temperature would permit to
assess the relevance of the {\it e-ph} interactions in the SSH
model. This paper addresses precisely this issue focussing on the
computation of some equilibrium thermodynamical properties. The
path integral method for the model Hamiltonian is outlined in
Section II while the results are given in Section III. Section IV
contains some final remarks.

\section*{II. Hamiltonian Model}

In one dimension the SSH Hamiltonian is:

\begin{eqnarray}
H=\,& & \sum_{r}J_{r,r+1} \bigl(f^{\dag}_r f_{r+1} +
f^{\dag}_{r+1} f_{r} \bigr) \,
 \nonumber \\
J_{r,r+1}=\,& & - {1 \over 2}\bigl[ J + \alpha (u_r -
u_{r+1})\bigr] \label{1}
\end{eqnarray}

where $J$ is the nearest neighbors hopping integral for an
undistorted chain, $\alpha$ is the electron-phonon coupling, $u_r$
is the displacement of the atomic group on the $r-$ lattice site
along the molecular axis, $f^{\dag}_r$ and $f_{r}$ create and
destroy electrons (i.e., $\pi$ band electrons in polyacetylene) on
the $r-$ group. The non-interacting Hamiltonian is given by a set
of independent oscillators. We transform the real space
Hamiltonian of (1) into a time dependent Hamiltonian
$H(\tau,\tau')$ by introducing $x(\tau)$ and $y(\tau')$ as the
electron coordinates at the $r$ and $r+1$ lattice sites,
respectively. The spacial {\it e-ph} correlations contained in (1)
are then mapped onto the time axis by changing: $ u_r \to u(\tau)$
and $u_{r+1} \to u(\tau')$. Accordingly we get:

\begin{eqnarray}
H(\tau,\tau')=\,& &  J_{\tau, \tau'} \Bigl(f^{\dag}(x(\tau))
f(y(\tau')) + f^{\dag}(y(\tau')) f(x(\tau)) \Bigr)\, \nonumber \\
J_{\tau, \tau'}=\,& & - {1 \over 2}\bigl[J + \alpha(u(\tau) -
u(\tau'))\bigr] \label{2}
\end{eqnarray}

Eq.(2) shows the semiclassical nature of the model in which
quantum mechanical degrees of freedom interact with the classical
variables $u(\tau)$.  As the electron hops are not constrained to
first neighbors sites along the chain $H(\tau,\tau')$ is more
general than the real space SSH Hamiltonian in (1). Setting
$\tau'=\,0$, $u(0)\equiv y(0) \equiv 0$, averaging the electron
operators over the ground state we obtain the average energy per
lattice site linearly depending on the displacements:

\begin{eqnarray}
{{<H(\tau)>} \over N}=& &\, V\bigl(x(\tau)\bigr) + u(\tau)j(\tau)
\nonumber \\ j(\tau)=& &\,-\alpha \Bigl(G[-x(\tau), -\tau ] +
G[x(\tau), \tau ]\Bigr) \label{3}
\end{eqnarray}

where $N=\,L/a$, with $L$ the chain length and $a$ the lattice
constant.  $V\bigl(x(\tau)\bigr)$ (proportional to $J$) is the
effective term accounting for the $\tau$ dependent electronic
hopping while $j(\tau)$ is the external source current for the
oscillator field, $G[x(\tau), \tau ]$ being the electron
propagator.

Taking a large number of oscillators ($u_i(\tau), i=1..{\bar N}$)
as the {\it bath} for the quantum mechanical particle whose
coordinate is $x(\tau)$, the general electron path integral  is
given by:

\begin{eqnarray}
& &<x(\beta)|x(0)>=\,\prod_i \int Du_i(\tau) \int Dx(\tau) \,
\nonumber \\  \times & &exp\Biggl[- \int_0^{\beta} d\tau \sum_i
{{M_i} \over 2} \Bigl(\dot{u_i}^2(\tau) + \omega_i^2 u_i^2(\tau)
\Bigr) \Biggr]  \, \nonumber \\ \times &
&exp\Biggl[-\int_0^{\beta} d\tau \biggl({m \over 2}
\dot{x}^2(\tau) + V\bigl(x(\tau)\bigr) - \sum_i
u_i(\tau)j(\tau)\biggr) \Biggr] \, \nonumber
\\ \label{4}
\end{eqnarray}

$\beta$ is the inverse temperature, $m$ is the electron mass and
$\omega_i$ are the oscillators frequencies. The oscillator masses
are considered as independent of $i$, $M_i\equiv M$ and hereafter
we set $M=\,10^4m$. After integrating out the oscillators
coordinates over the paths $Du_i(\tau)$ and imposing a closure
condition $\Bigl(x(\beta)=\,x(0)\Bigr)$ on the particle paths, we
obtain the total partition function in the functional form:

\begin{eqnarray}
& &Z(j(\tau))=\,Z_{ph} \oint Dx(\tau)exp\Bigl[-{m \over 2}
\dot{x}^2(\tau) - V\bigl(x(\tau)\bigr) - A(j(\tau)) \Bigr] \,
\nonumber \\ & &Z_{ph}=\, \prod_{i=1}^{\bar N} {1 \over
{2\sinh(\hbar\omega_i\beta/2)}} \, \nonumber \\ & &A(j(\tau))=\,
-{{\hbar^2} \over {4M}}\sum_{i=1}^{\bar N} {1 \over {\hbar
\omega_i \sinh(\hbar\omega_i\beta/2)}} \, \nonumber \\ \times & &
\int_0^{\beta} d\tau j(\tau) \int_0^{\beta}d{\tau''}
\cosh\Bigl(\omega_i \bigl( |\tau - {\tau''}| - \beta/2 \bigr)
\Bigr) j({\tau''}) \, \nonumber
\\
\label{5}
\end{eqnarray}

The nonequilibrium quantum statistics of the system can be derived
via (4) through the closed-time path formalism \cite{schwinger}
which permits to evaluate dissipative properties due to the phonon
bath friction \cite{haba}. The thermodynamical properties of the
system follow from (5).

\section*{III. Phonon Anharmonicity}

The electron particle path interacts with each of the ${\bar N}$
oscillators through the coupling $\alpha$ (assumed independent of
$i$) of the SSH Hamiltonian. Then, the $x(\tau)-$ perturbed phonon
partition function can be expanded in anharmonic series as:

\begin{eqnarray}
Z_{ph}[x(\tau)]\simeq \,& & Z_{ph} \biggl(1 + \sum_{l=1}^k (-1)^l
<C^l>_x \biggr)\, \nonumber \\ <C^k>_{x}=\, & &Z_{ph}^{-1} \prod_i
\oint Du_i(\tau) {1 \over {k!}} \prod_{l=1}^k \Biggl[
\int_0^{\beta} d\tau_l u_i(\tau_l) j(\tau_l) \Biggr]^{l}  \,
\nonumber \\ \times & & exp\Biggl[- \int_0^{\beta} d\tau \sum_i
{M_i \over 2} \bigl( \dot{u_i}^2(\tau) + \omega_i^2 u_i^2(\tau)
\bigr) \Biggr]\, \nonumber \\ \label{6}
\end{eqnarray}

The total phonon partition function $Z_{ph}^T$ is obtained by
integrating (6) over a set of perturbing electron paths. Since the
oscillators are decoupled (and anharmonic effects mediated by the
electron particle path are here neglected) we can study the
behavior of the cumulant terms $<C^k>_{x}$ by selecting a single
oscillator having energy $\omega$ and displacement $u(\tau)$.

As a first step let's replace the source current of the SSH model
by a linear (in the electron path) current $j_x(\tau)=\,  -\alpha
x(\tau)$. This allows us to derive analytical expression for the
cumulants as shown, for the lowest order, in the Appendix. Let's
approximate the electron path by its $\tau$ averaged value:
 $<x(\tau)>\equiv\,{1 \over \beta}\int_0^{\beta} d\tau x(\tau) =\,x_0/a$.
and expand the oscillator path in $N_F$ Fourier components:

\begin{eqnarray}
& &u(\tau)=\,u_o + \sum_{n=1}^{N_F} 2\Bigl(\Re u_n \cos( \omega_n
\tau) - \Im x_n \sin( \omega_n \tau) \Bigr)\, \nonumber \\ &
&\omega_n=\,2\pi n/\beta \label{7}
\end{eqnarray}

Taking the  measure of integration

\begin{eqnarray}
\oint Du(\tau)\equiv  & &\biggl({1 \over 2}\biggr)^{2N_F} {{\Bigl(
2\pi \cdot \cdot 2N_F \pi \Bigr)^2} \over {\sqrt{2}
\lambda_M^{(2N_F+1)} }} \int_{-\infty}^{\infty}{du_o} \, \nonumber
\\  \times & &\prod_{n=1}^{N_F} \int_{-\infty}^{\infty} d\Re u_n
\int_{-\infty}^{\infty} d\Im u_n \label{8}
\end{eqnarray}

with $\lambda_M=\,\sqrt{\pi \hbar^2 \beta/M}$, we obtain for the
$k-th$ cumulant the expression

\begin{eqnarray}
& &<C^k>_{N_F}=\,Z^{-1}_{ph} {{(\alpha_{R} \beta \lambda_M)^k (k -
1)!!} \over {k! \pi^{k/2} (\omega \beta)^{k+1}}} \prod_{n=1}^{N_F}
{{(2n\pi)^2} \over {(2n\pi)^2 + (\omega \beta)^2}}\, \nonumber \\
& &\alpha_R=\,\alpha x_0/a \label{9}
\end{eqnarray}

Let's set $x_0/a=\,0.1$ in the following calculations thus
reducing the effective coupling $\alpha_R$  by one order of
magnitude with respect to the bare value. The trend shown by the
results hereafter presented does not depend however on this choice
since $x_0/a$ and $\alpha$ can be varied independently. As the
cumulants should be stable against the number of Fourier
components in the oscillator path expansion, using (9) we set the
minimum $N_F$ through the condition $2N_F \pi \gg \omega \beta$.
Noticing that odd$-k$ terms vanish the thermodynamics of the
anharmonic oscillator can be computed by the cumulant corrections
to the harmonic phonon free energy:

\begin{equation}
F^{(k)}=\,-{1 \over \beta} \ln \Bigl[1 +
\sum_{l=1}^{k}<C^{2l}_{N_F}> \Bigr] \label{10}
\end{equation}

To proceed I need a criterion to find the temperature dependent
cutoff $k^*$ in the cumulant series.  In the low $T$ limit the
third law of thermodynamics may offer the suitable constraint to
determine $k^*$. Then, given $\alpha$ and $\omega$, the program
searches for the cumulant order such that the heat capacity and
the entropy tend to zero in the zero temperature limit. At any
finite temperature T, the constant volume heat capacity is
computed as

\begin{eqnarray}
C_V^{(k)}(T)=\,& &- \Bigl[F^{(k)}(T + 2\Delta) - 2F^{(k)}(T +
\Delta) + F^{(k)}(T) \Bigr] \nonumber \\  \times & &\Bigl({1 \over
\Delta} + {T \over \Delta^2} \Bigr) \label{11}
\end{eqnarray}

$\Delta$ being the incremental step and $k^*$ is determined as the
minimum value for which the heat capacity converges with an
accuracy of $10^{-4}$. Figs.1(a) and 1(b) show phonon heat
capacity and free energy respectively in the case of a low energy
oscillator for an intermediate value of {\it e-ph} coupling.
Harmonic functions, anharmonic functions with second order
cumulant and anharmonic functions with $k^*$ corrections are
reported on in each figure. The second order cumulant is clearly
inadequate to account for the low temperature trend yielding a
negative phonon heat capacity below $\sim 40K$ while at high $T$
the second order cumulant contribution tends to vanish. Instead,
the inclusion of $k^*$ terms in eqs.(10) and (11) leads to the
correct zero temperature limit although there is no visible
anharmonic effect on the phonon heat capacity throughout the whole
temperature range being $C_V^{(k^*)}$ perfectly superimposed to
the harmonic $C_V^h$. Note in Fig.1(b) that the $k^*$ corrections
simply shift downwards the free energy without changing its slope
versus temperature. By increasing $\alpha_R$, the low T range with
wrong (negative) $C_V^{(2)}$ broadens (see Fig.2(a)) whereas the
$k^*$ contributions permit to fulfill the zero temperature
constraint and substantially lower the phonon free energy (see
Fig.2(b)). Thus, for the particular choice of constant (in $\tau$)
source current we find that the {\it e-ph} anharmonicity
renormalizes the phonon partition function although no change
occurs in the thermodynamical behavior of the free energy
derivatives. Anharmonicity is essential to stabilize the system
but it leaves no trace in the heat capacity. Incidentally we note
that a similar role for the anharmonic interactions has been
proposed to explain the low frequency behavior of glasses
\cite{gurevich}. Figure 3(a) displays the $k^*$ temperature
dependence for three choices of {\it e-ph} coupling in the case of
a low energy oscillator: while, at high $T$, the number of
required cumulants ranges between six and ten according to the
coupling, $k^*$ strongly grows at low temperatures getting the
value 100 at $T=\,1K$ for $\alpha_R=\,60meV \AA^{-1}$. The $k^*$
versus $\alpha_R$ behavior is pointed out in Fig.3(b) for three
selected temperatures. Figure 3(c) shows that by decreasing the
oscillator energy an increasing number of cumulants has to be
taken into account to make the thermodynamical functions
convergent: this is consistent with the expectations that high
energy oscillators are less sensitive to {\it e-ph} induced
anharmonicity.

Next we turn to the computation of the equilibrium thermodynamics
of the phonon subsystem in our semiclassical SSH model hence,
using in (6) the external source current given in (3). As the
electron propagator depends on the bare hopping integral we set
$J=100meV$ thus assuming a narrow band electron system. Any
electron path yields in principle a different cumulant
contribution as pointed out in the Appendix. Numerical
investigation shows however that convergent $k-$ order cumulants
are achieved by: i) taking $M_F=\,2$ Fourier components in the
electron path expansion, ii) setting for the coefficients $\{x_o,
c_m\}$ a maximum amplitude of order 0.1 (in units of the lattice
constant) and iii) summing over $\sim 5^{2M_F+1}$ electron paths.

As in the previous case, we truncate the cumulant series in (6) by
invoking the third law of thermodynamics to determine the cutoff
$k^*$ in the low temperature limit and by searching numerical
convergence on the first and second free energy derivatives at any
finite temperature. $k^*$ does not depend on the specific electron
path coefficients then,
$<C^{k^*}>=\,\sum_{(x_o,c_m)}<C^{k^*}>_{(x_o,c_m)}$. Again, we can
start our analysis from (10) after checking that odd $k$ cumulants
yield vanishing contributions. Now however the picture of the
anharmonic effects changes drastically. The {\it e-ph} coupling
strongly modifies the shape of the heat capacity and free energy
plots with respect to the harmonic result as it is seen in
Figs.4(a) and 4(b) respectively. Above a threshold value $\alpha
\sim 10meV \AA^{-1}$ (depending on the energy oscillator here set
at $\omega=\,20meV$) the heat capacity versus temperature curves
show a peak whose location along the $T$ axis is reported on in
Fig.4(d) as a function of $\alpha$. By enhancing $\alpha$ the
height of the peak grows and the bulk of the anharmonic effects on
the heat capacity is shifted towards lower $T$. At $\alpha \sim
60meV \AA^{-1}$ the crossover temperature is around 100K. At high
$T$ the anharmonic corrections renormalize downwards the free
energy but their effect on the heat capacity tends to decrease
signalling that {\it e-ph} nonlinearities are rather to be seen in
the intermediate to low temperature range.

Also the $k^*$ versus $T$ behavior is much different from the
previous case (discussed in Fig.3(a)): it is seen in Fig.4(c) that
a few cumulant terms suffice at low temperatures even at large
{\it e-ph} couplings while $k^*$ grows by increasing $T$ for any
$\alpha$. Then, $k^*(T)$ does not necessarily provide a measure of
the degree of {\it e-ph} anharmonicity as revealed by the heat
capacity behavior. At room temperature we find $k^*=\,14$ as the
maximum value for the range of $\alpha$'s here considered with a
low energy oscillator. However, by taking $\omega=\,100meV$, $k^*$
drops (as shown in Fig.4(c)) attaining the value 6 at around room
temperature for $\alpha=\,60meV \AA^{-1}$. This would suggest that
also in the SSH model high energy phonons are expected to be less
anharmonic. This idea is confirmed by looking at Fig.4(d) where
the crossover temperature in the oscillator heat capacity curves
is reported on, at fixed $\alpha$, versus $\omega$. The heat
capacity peak shows up at $\omega \leq 50meV$ and its temperature
location is shifted downwards by decreasing the phonon energy
whereas, at larger energies, the heat capacity has a positive
derivative in the whole range of temperatures. Finally we note
that recent numerical studies of a classical one dimensional
anharmonic model undergoing a Peierls instability \cite{allen}
also find a specific heat peak as a signature of anharmonicity.

\section*{IV. Conclusions}

We have studied the equilibrium thermodynamics of an
electron-phonon system looking in particular at the anharmonic
effects produced by the electronic subsystem on the phonon
oscillators. The path integral formalism permits to analyse the
{\it e-ph} non linearities as a function of the source current
(peculiar of the Hamiltonian model) which disturbs the harmonic
oscillator. The phonon partition function has been expanded in
$k-$ order cumulant series and, at any temperature, we have
determined the cutoff $k^*$ that makes first and second free
energy derivatives convergent. In the zero temperature limit the
constraint imposed by the third law of thermodynamics has been
fulfilled. Rather than being a unique measure of {\it e-ph}
anharmonicity $k^*(T)$ turns out to be a model dependent function
whose  values may vary considerably according to the physical
quantities one decides to sample. As a general trend we find
however that, at fixed $T$ and {\it e-ph} coupling, higher energy
oscillators are more stable and require a lower $k^*$ for the
computation of their thermodynamical properties.

We have first considered the case of a source current proportional
to the time averaged electron path and analytically derived the
cumulant expansion for the phonon partition function. This current
induces a slope preserving shift in the phonon free energy versus
temperature hence, the heat capacity does not show any correction
with respect to the harmonic result in spite of the high number of
anharmonic terms which appear in the cumulant series.

Turning to the source current peculiar of the time dependent, one
dimensional, semiclassical Su-Schrieffer-Heeger model we find a
striking evidence of {\it e-ph} anharmonicity in the constant
volume heat capacity of a single oscillator. As a main feature the
heat capacity exhibits a peak whose height and location on the $T$
axis varies with the strength of the {\it e-ph} coupling and the
energy of the oscillator: while strong couplings shift the main
body of the anharmonic effects towards low $T$, high energy
phonons prove to be less affected by {\it e-ph} corrections.

\begin{figure}
\vspace*{10truecm} \caption{(a) Phonon heat capacity and (b)
Phonon free energy calculated in i) the harmonic model, ii)
anharmonic model with second order cumulant, iii) anharmonic model
with $k^*$ cumulants (see text).  $\alpha_R$ is the effective {\it
e-ph} coupling in units $meV \AA^{-1}$ and $\omega$ is the phonon
energy.}
\end{figure}

\begin{figure}
\vspace*{10truecm} \caption{ As in Fig.1 but with larger {\it
e-ph} coupling.}
\end{figure}

\begin{figure}
\vspace*{10truecm} \caption{(a) Number of cumulants required to
obtain a convergent phonon heat capacity at any temperature for
different choices of {\it e-ph} couplings. (b) Number of cumulants
yielding a convergent phonon heat capacity at any {\it e-ph}
coupling for three selected temperatures. (c) Number of cumulants
yielding a convergent phonon heat capacity at any oscillator
energy for three selected temperatures.}
\end{figure}

\begin{figure}
\vspace*{10truecm} \caption{ Anharmonic (a) phonon heat capacity
and (b) free energy versus temperature for eight values of {\it
e-ph} coupling. The harmonic plots are also reported on. A low
energy oscillator is assumed.  (c) Temperature dependence of the
cutoff $k^*$ for: i) the weakest and strongest {\it e-ph}
couplings given in (a) and (b) with $\omega=20meV$, ii) the
strongest coupling case with $\omega=100meV$. (d) Temperatures at
which the heat capacity gets the maximum value (see (a)) versus
{\it e-ph} coupling at fixed $\omega$ (bottom axes) and versus
$\omega$ at fixed coupling (top axes).}
\end{figure}

\section*{Appendix}

Let's assume a source current $j_x(\tau)=\,  -\alpha x(\tau)$ as
the disturbing term for the oscillator field, $\alpha$ being the
{\it e-ph} coupling and $x(\tau)$ being the electron path. Then
the {\it k-th} order cumulant term contributing to the $x(\tau)-$
perturbed phonon partition function is given by

\begin{eqnarray}
<C^k>_{x}=& &\,Z_{ph}^{-1}  {{\alpha^k} \over {k!}} \oint Du(\tau)
\prod_{l=1}^k \Biggl[\int_0^{\beta} d\tau_l
u(\tau_l)x(\tau_l)\Biggr]^{l} \cdot \,\nonumber \\ & & exp\Biggl[-
\int_0^{\beta} d\tau \sum_i {M_i \over 2} \bigl( \dot{u}^2(\tau) +
\omega^2 u^2(\tau) \bigr) \Biggr] \label{12}
\end{eqnarray}

We use the oscillator path and the functional measure of
integration given in the eqs. (7) and (8) of the text
respectively. Then, the electron path expansion in Fourier
components

\begin{eqnarray}
& &x(\tau)=\,x_o + \sum_{m=1}^{M_F} \Bigl(c_m \cos( \omega_m \tau)
+ d_m \sin( \omega_m \tau) \Bigr) \, \nonumber \\ &
&\omega_m=\,2\pi m/\beta \, \nonumber \\ & & c_m=\, 2 \Re x_m \,
\nonumber \\ & & d_m=\, -2 \Im x_m \label{13}
\end{eqnarray}

will be truncated at $M_F=\,N_F$. Without any loss of generality
we set $d_m=\,0$ and derive the following expressions for the
lowest even order cumulants:

\begin{eqnarray}
<C^2>& &_{(x_o,c_m)}=\,Z_{ph}^{-1}{\alpha^2 \over 2}{(\beta
\lambda_M)^2 \over {\beta \omega}} F(N_F) \cdot \Biggl[ {{x_o^2}
\over {\pi (\beta \omega)^2}} + \nonumber
\\ & &{1 \over
{2\pi}}\sum_{m=1}^{N_F} {{ c_m^2} \over { (2\pi m)^2 + (\beta
\omega)^2 }} \Biggr] \, \nonumber
\\
<C^4>& &_{(x_o,c_m)}=\,Z_{ph}^{-1}{\alpha^4 \over {4!}}{(\beta
\lambda_M)^4 \over {\beta \omega}} F(N_F) \cdot \Biggl[ {{3x_o^4}
\over {2 \pi^2 (\beta \omega)^4}} + \nonumber
\\ & &{{3x_o^2} \over {2 \pi^2
(\beta \omega)^2}} \sum_{m=1}^{N_F}  {{2 c_m^2} \over {(2\pi m)^2
+ (\beta \omega)^2 }} +  \, \nonumber
\\ & &{3 \over {8 \pi^2}} \sum_{{{m=1}, {l>m}}}^{N_F}
{{2 c_m^2} \over {(2\pi m)^2 + (\beta \omega)^2}}\cdot {{2 c_l^2}
\over {(2\pi l)^2 + (\beta \omega)^2}}  + \nonumber \\ & &{3 \over
{4\pi^2}} \sum_{m=1}^{N_F} {{ c_m^4} \over { ((2\pi m)^2 + (\beta
\omega)^2)^2}} \Biggr]\, \nonumber
\\
F(N_F)=& &\,\prod_{m=1}^{N_F} {{(2\pi m)^2} \over {(2\pi m)^2 +
(\beta \omega)^2}} \label{14}
\end{eqnarray}

Odd $k-$ cumulants vanish at any order. Setting a maximum
amplitude for the coefficients $x_o$ and $c_m$ and integrating
eqs.(14) over a class of electron paths one finds the total
cumulant contributions to the partition function.

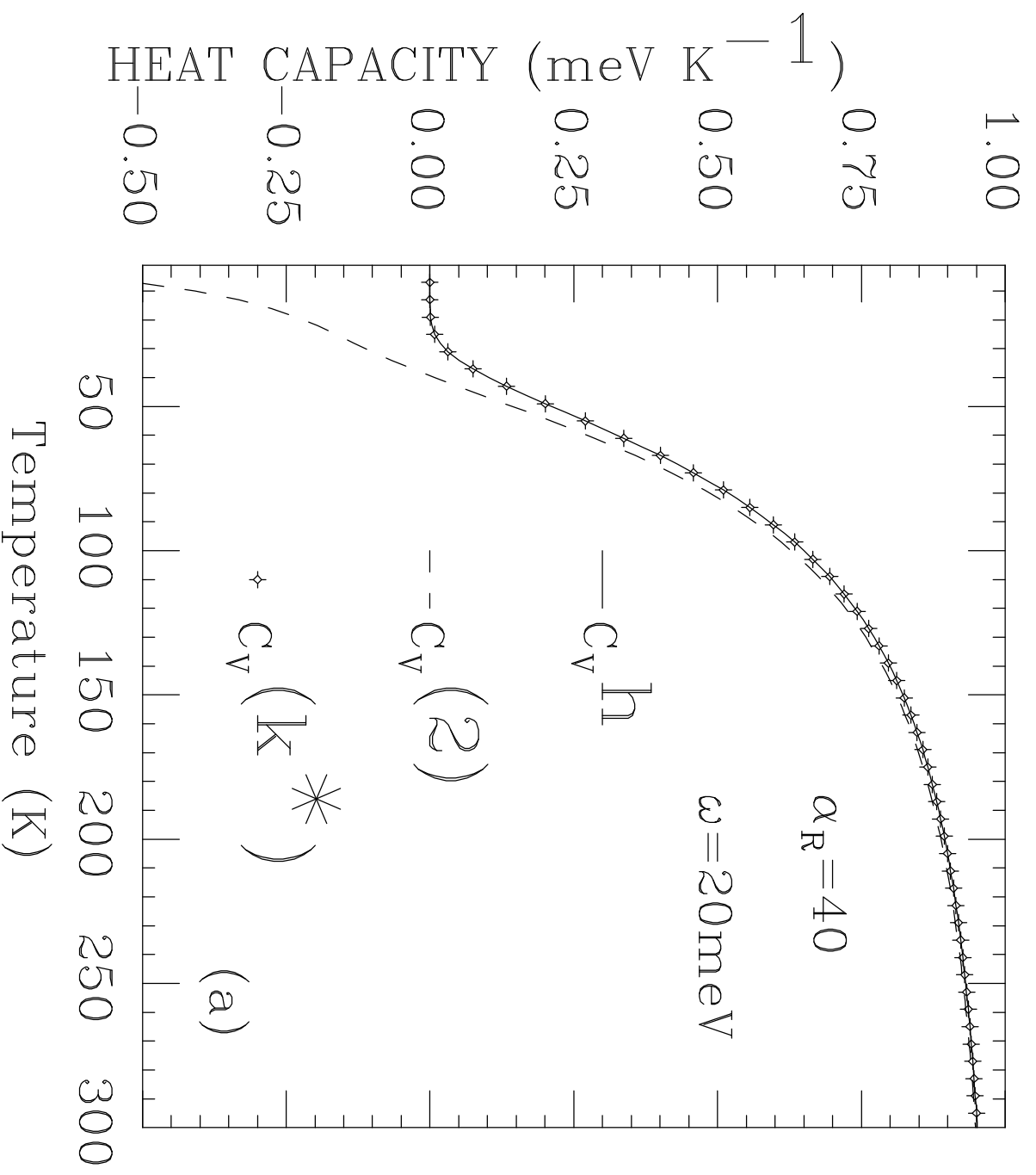
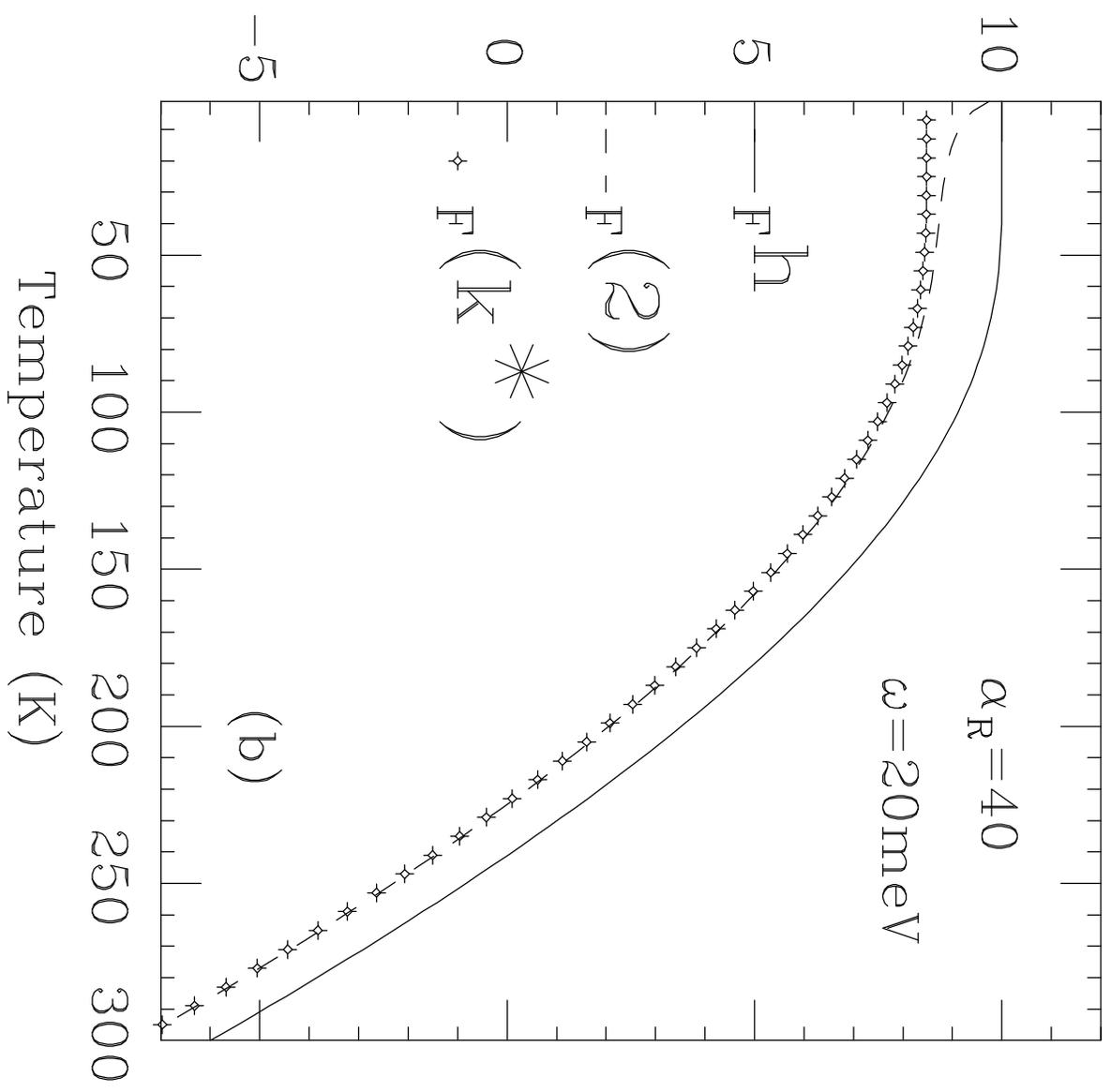
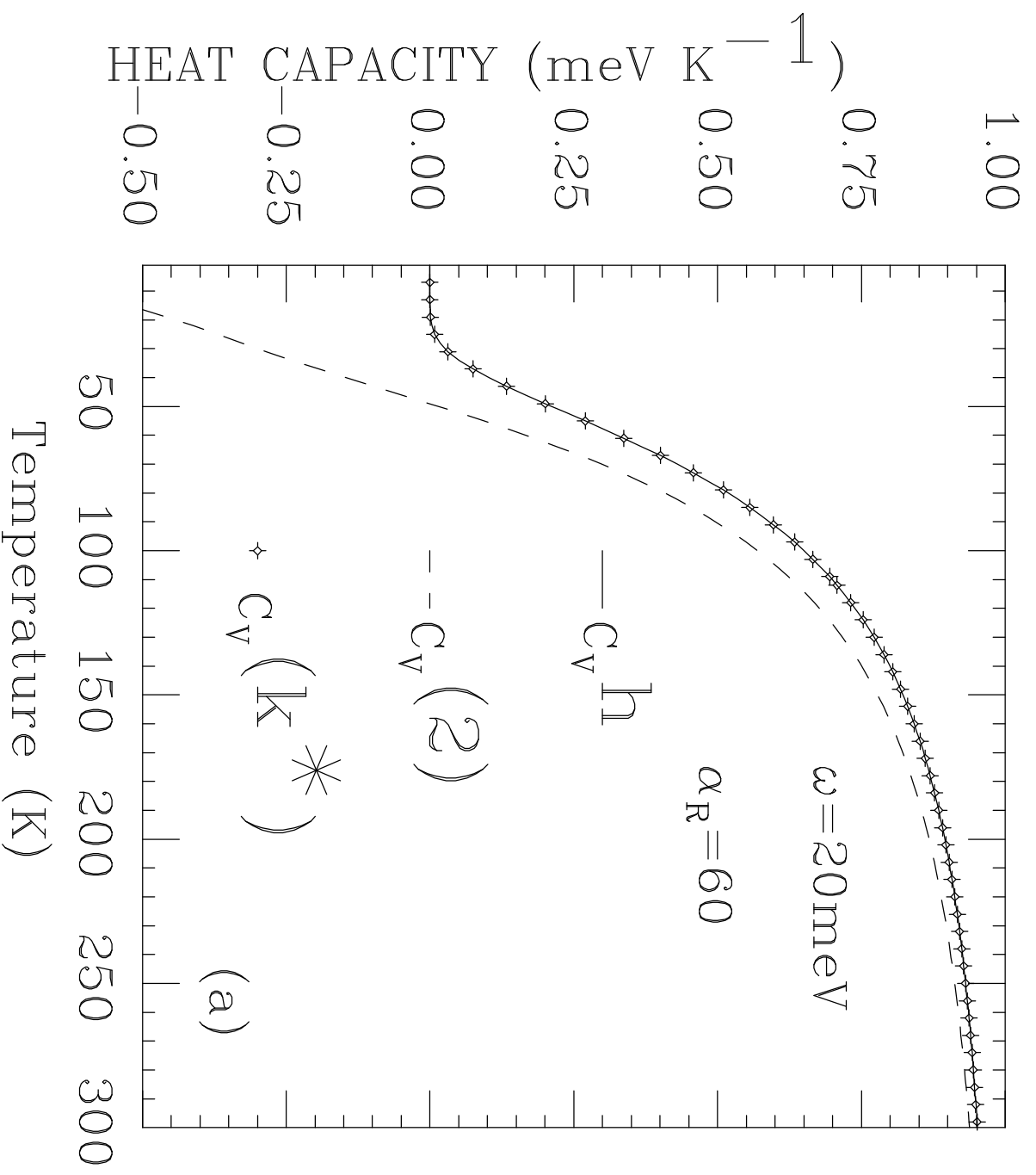
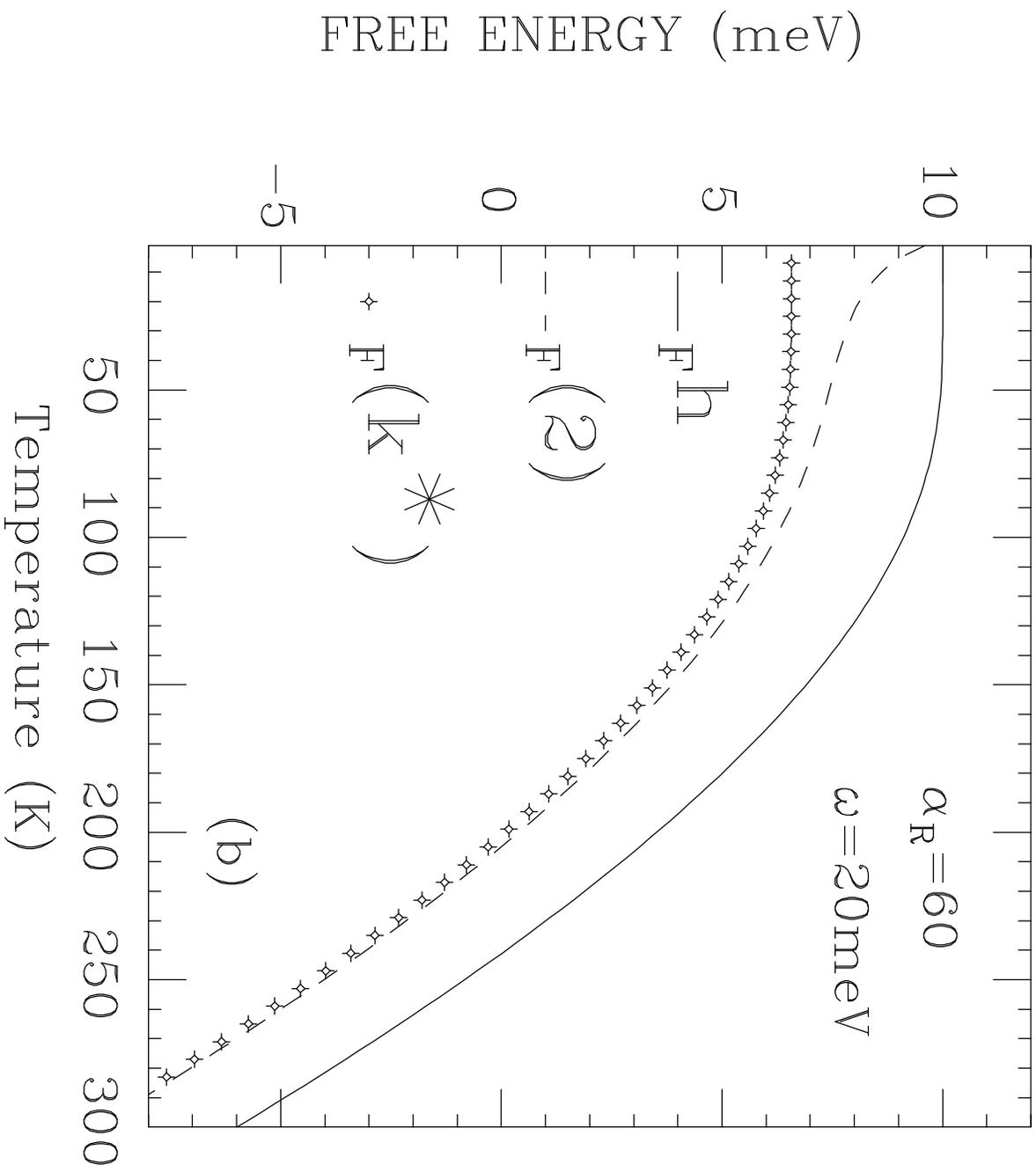
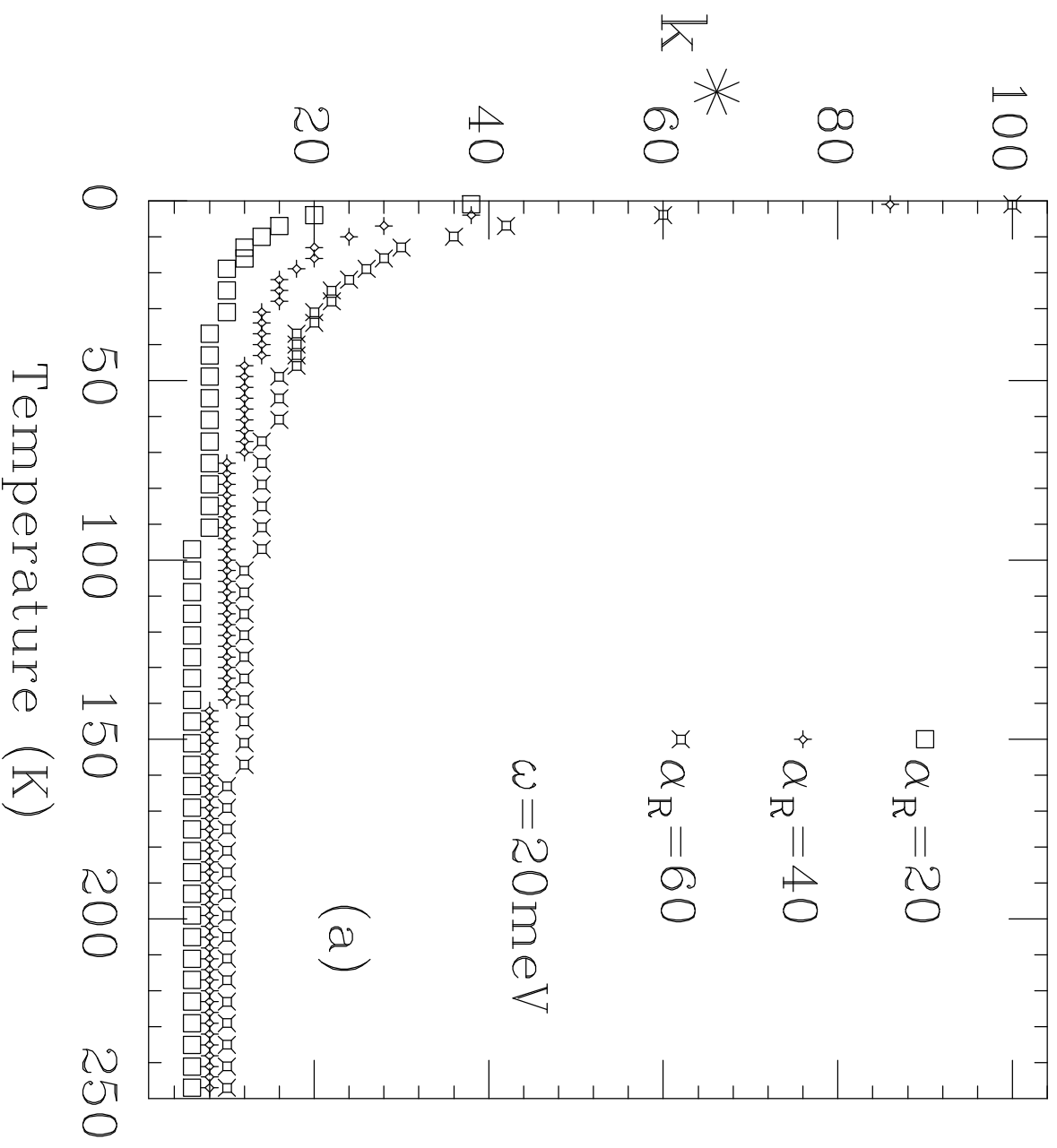
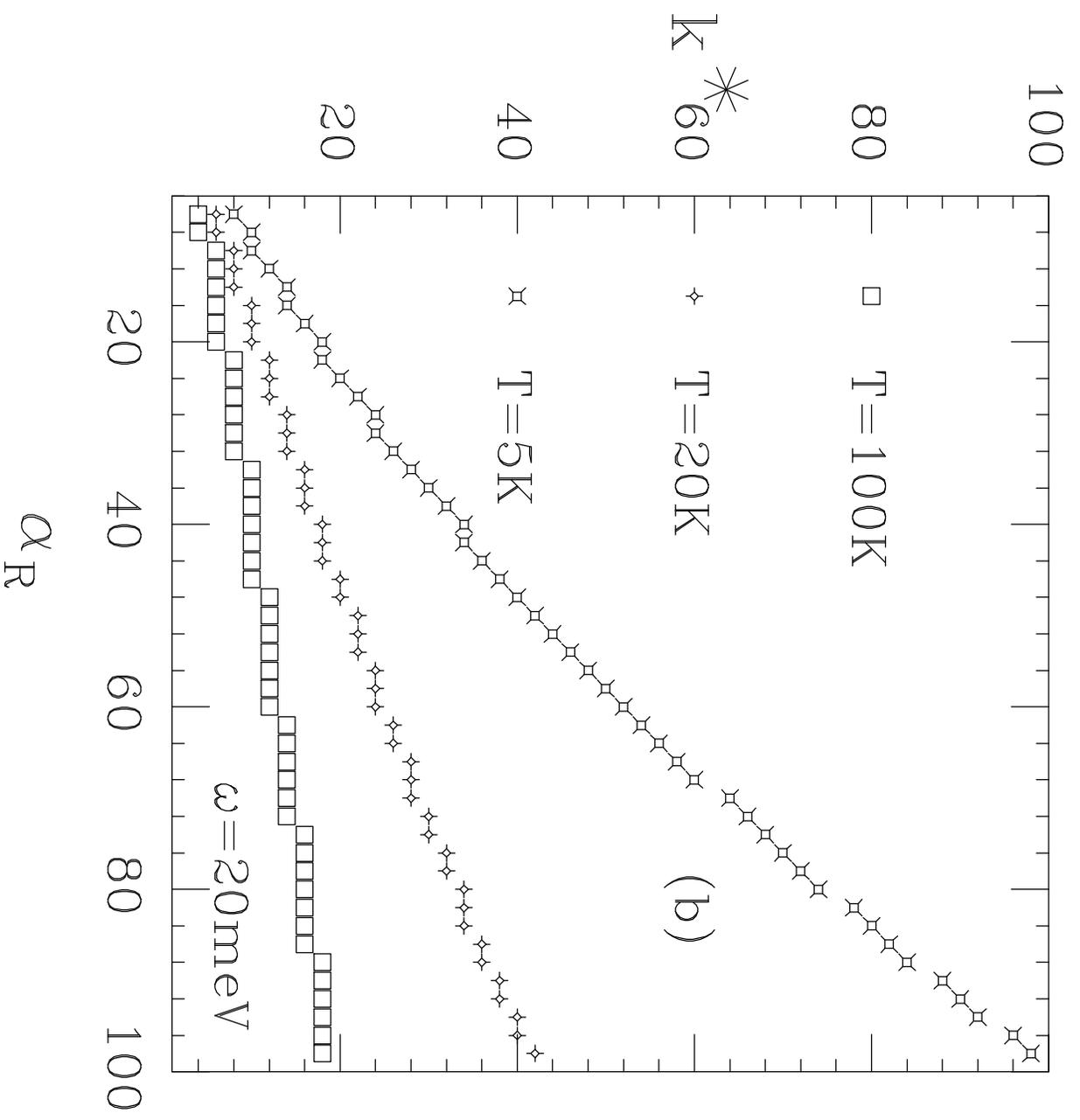
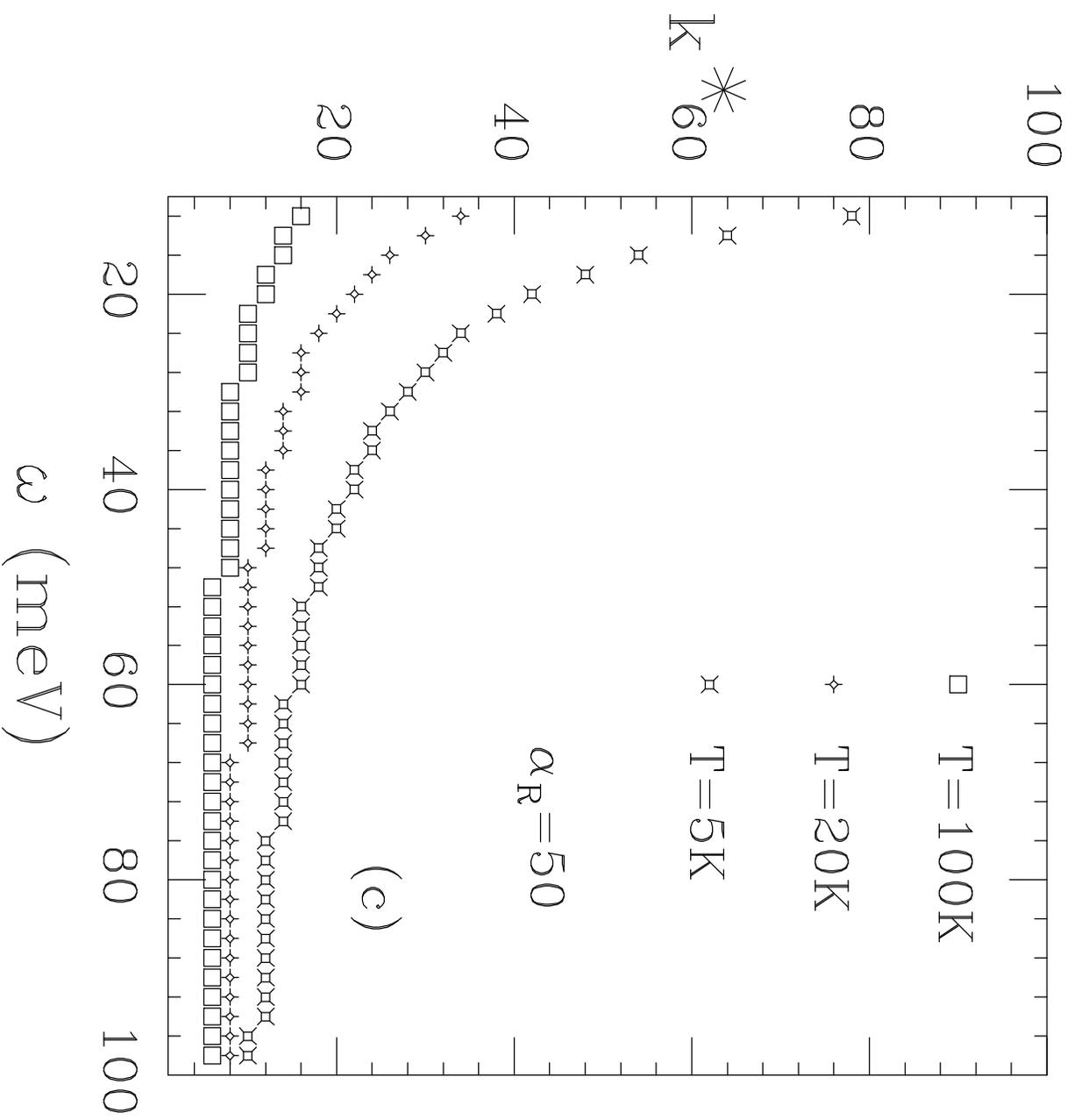
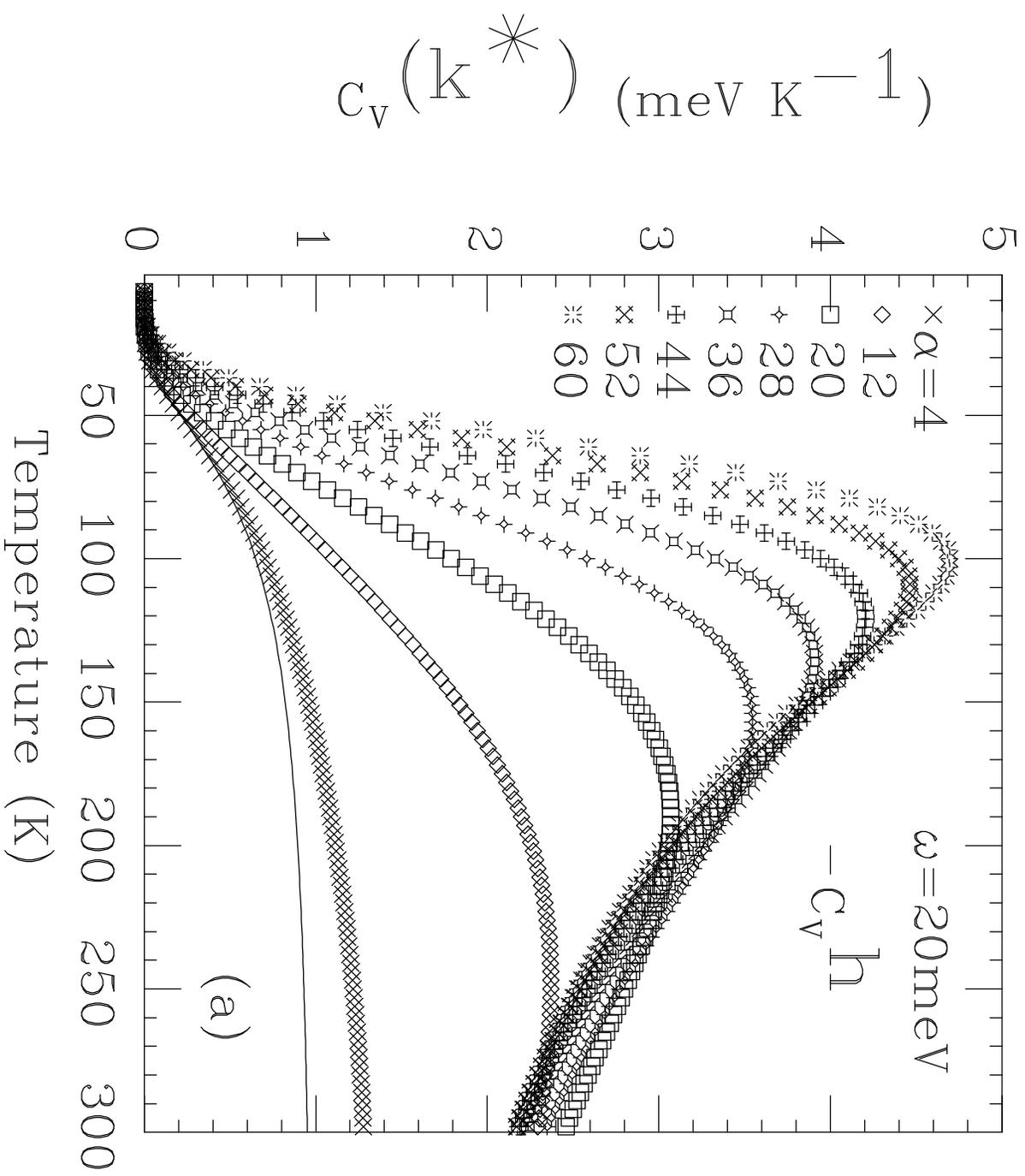
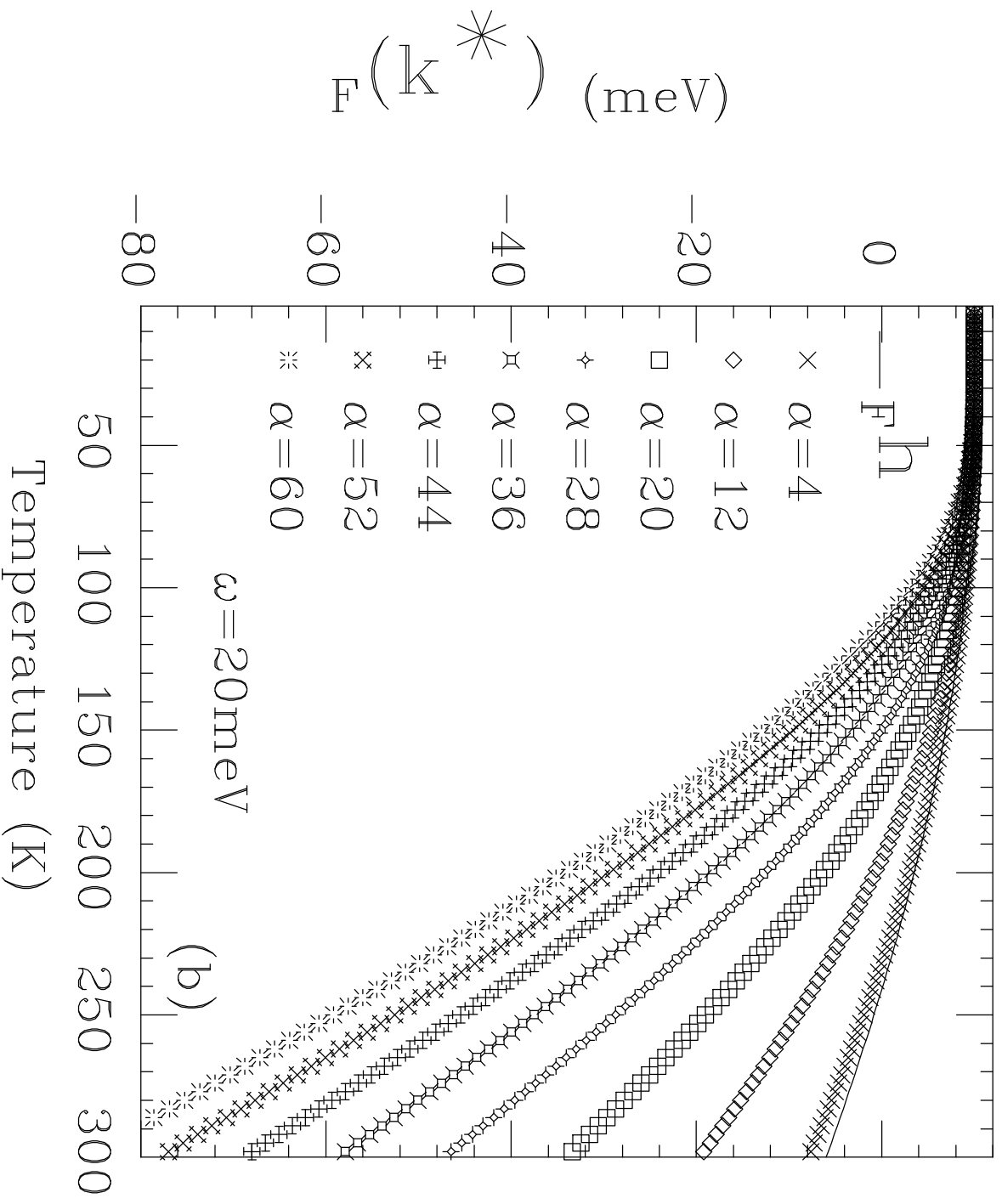
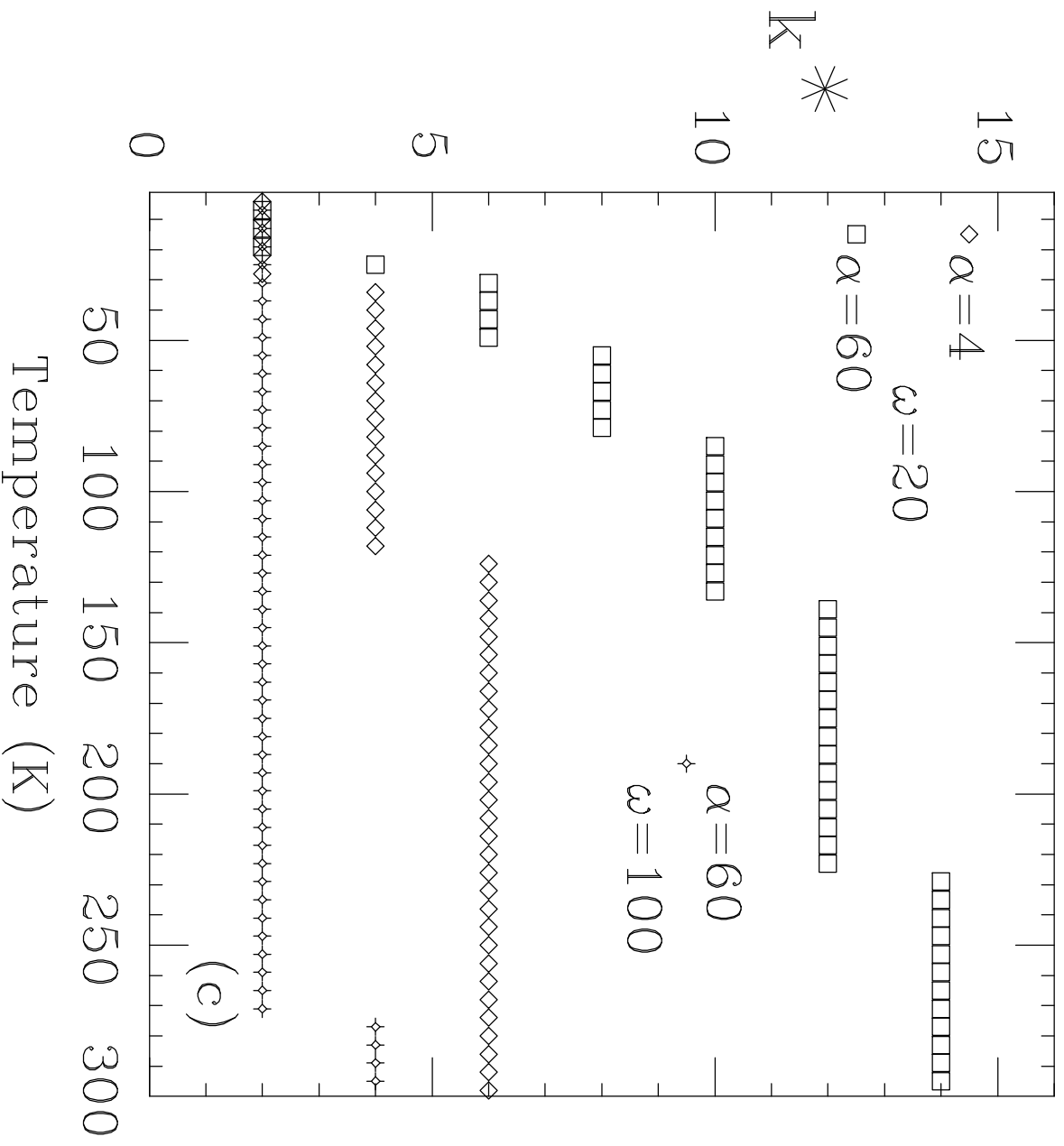
\end{document}